\documentclass[]{CUP-JNL-DCE}

\usepackage[utf8]{inputenc}
\usepackage{latexsym}
\usepackage{graphicx}
\usepackage{multicol,multirow}
\usepackage{amsmath,amssymb,amsfonts}
\usepackage{mathrsfs}
\usepackage{amsthm}
\usepackage{apacite}
\usepackage{rotating}
\usepackage{appendix}
\usepackage[authoryear]{natbib}
\usepackage{hyperref}
\usepackage{cleveref}
\usepackage{booktabs}
\usepackage{tabularx}
\usepackage{array}
\usepackage{float}
\usepackage{siunitx}

\usepackage{caption}
\usepackage{subcaption}
\usepackage{graphbox}
\usepackage{enumitem}
\usepackage{makecell}

\newcommand*\diff{\mathop{}\!\mathrm{d}}

\articletype{RESEARCH ARTICLE}
\jname{Data-Centric Engineering}
\jyear{2020}

\begin{document}

\begin{Frontmatter}

\title[Constructing Digital Twins for Manufacturing]
{Numerical simulation, clustering and prediction of multi-component polymer precipitation}

\author[1,2]{Pavan Inguva}
\author[3]{Lachlan Mason}
\author[3,4]{Indranil Pan}
\author[2]{Miselle Hengardi}
\author*[2]{Omar K. Matar}\email{o.matar@imperial.ac.uk}

\authormark{Pavan Inguva \textit{et al.}}

\address[1]{\orgdiv{Department of Chemical Engineering}, \orgname{Massachusetts Institute of Technology}, \orgaddress{ \postcode{Cambridge 02139}, \country{United States}}}

\address[2]{\orgdiv{Department of Chemical Engineering}, \orgname{Imperial College London}, \orgaddress{ \postcode{London SW7 2AZ}, \country{United Kingdom}}}

\address*[3]{\orgdiv{Data Centric Engineering Program}, \orgname{The Alan Turing Institute}, \orgaddress{\street{96 Euston Road}, \postcode{London NW1 2DB}, \country{United Kingdom}}}

\address*[4]{\orgdiv{Centre
for Environmental Policy}, \orgname{Imperial College London}, \orgaddress{\postcode{London SW7 2AZ}, \country{United Kingdom}}}



\abstract{
Multi-component polymer systems are of interest in organic photovoltaic and drug delivery applications, among others where diverse morphologies influence performance. An improved understanding of morphology classification, driven by composition-informed prediction tools, will aid polymer engineering practice.
We use a modified Cahn–Hilliard model to simulate polymer precipitation. Such physics-based models require high-performance computations that prevent rapid prototyping and iteration in engineering settings. To reduce the required computational costs, we apply machine learning techniques for clustering and consequent prediction of the simulated polymer blend images in conjunction with simulations. Integrating ML and simulations in such a manner reduces the number of simulations needed to map out the morphology of polymer blends as a function of input parameters and also generates a data set which can be used by others to this end. 
We explore dimensionality reduction, via principal component analysis and autoencoder techniques, and analyse the resulting morphology clusters.  Supervised machine learning using Gaussian process classification was subsequently used to predict morphology clusters according to species molar fraction and interaction parameter inputs. Manual pattern clustering yielded the best results, but machine learning techniques were able to predict the morphology of polymer blends with \SI{\geq 90}{\percent} accuracy.
}

\begin{policy}[Impact Statement]
By providing predictive tools to polymer engineers that assist in predicting morphological features from easily knowable (or measurable) input parameters, the need to perform time-consuming and costly experiments to obtain desired morphological behaviours in such complex systems is reduced. This could reduce the overall complexity of the Research and Development (R\&D) process for a variety of industries. Applying data-driven approaches to analysing simulation data may also help to identify trends and features that are important but might not otherwise be detected by humans.
\end{policy}

\end{Frontmatter}


\section{Introduction}
\label{sec1}

Multi-component polymer systems are of industrial interest in a variety of applications such as high-performance plastics \citep{Naum1994}, membrane systems\citep{Yang2018,Ulb2006}, nanoparticle and nano-colloidal systems \citep{Lee2017,Li2017}, and drug delivery \citep{Lao2008, Inguva2015}. One of the key features considered during the research and development (R\&D) and/or manufacturing process is the morphology of the polymeric particles/blends formed. The morphology can profoundly impact the final product's performance and usability as for many applications, there is an optimal morphology that is desired. Understanding the relationship between polymer properties and their resultant blend morphology will therefore help guide product development from synthesis to manufacturing steps.

Computational methods provide an excellent tool for modelling various phenomena across various scales in polymeric systems \citep{Goon2017}. Modelling techniques can be applied to understand and evaluate a variety of thermodynamic and transport properties such as polymer-blend miscibility. They can also be used in engineering and manufacturing applications to understand how morphological structures form or how materials respond to processing conditions.
%
On a molecular/atomic scale, techniques such as molecular dynamics have been widely used in polymer systems to evaluate properties such as the miscibility and interactions of polymers in a blend or with other species \citep{Luo2010,Prat2007}, diffusion coefficients and transport characteristics \citep{Pave2005}, composite elasticity \citep{Han2007}, and nanoparticle morphology \citep{Li2017}. Recent, non-equilibrium molecular dynamics simulation studies have also considered the effect of shear on the morphology of anisotropic nanoparticles \citep{Bian2015,Arau2016} such as Janus nanoparticles which are an interesting case within the possibility space of multi-component polymer systems.

Continuum-based techniques are typically applied at length and time scales orders of magnitude larger than discrete approaches. Phase-field models such as the Cahn–Hilliard equation \citep{Cahn1958} are useful in capturing the dynamic behaviour of structures and morphologies in heterogeneous systems. The Cahn–Hilliard model accounts for various thermodynamic driving forces for morphology evolution, such as homogeneous free energy and interfacial energy, and can also be adapted to further account for other relevant transport phenomena such as convection\citep{Wodo2012}.
Previous continuum-scale simulations of multi-component polymer systems have focused `uphill' diffusion, as described by the Cahn–Hilliard equation. Examples of previous applications include systems of two polymers and one solvent (denoted PPS) \citep{Shan2010,Alfa2007}, or ternary polymer systems \citep{Naum1994,Alfa2007}. Studies that have considered the influence of convective transport in multi-component systems have only evaluated a single polymer species precipitating out of solution \citep{Tree2017,Zhou2006}. Consequently, there is a knowledge gap in the understanding of how systems containing more than one polymer, i.e.~PPS and PPP systems, behave in the presence of convective mass transport.   

Machine learning (ML) has increasingly found use in physical simulations and computational modelling by complementing or replacing traditional modelling approaches. ML is noted in having strength in pattern recognition and data mining \citep{Brunton2020} which correspondingly enables it to be used for many tasks relevant to physical simulations. Previous studies have applied ML to develop data-driven surrogate/reduced order models (ROMs) to improve the speed of computations and results generation \citep{Peher2017, Janet2018, San2018}. ML has also been used to improve the accuracy of physical simulations by enabling the development of data-driven closures \citep{Durai2019} and models \citep{Chmiela2017} which can capture more data than traditional first principles or empirical models. Within the material sciences, ML has similarly found increasing use in a variety of cases ranging from the prediction of macroscopic self-assembled structures using molecular properties \citep{Inok2018} to the prediction of novel permanent magnets \citep{Moll2018} and in the optimisation of alloy properties \citep{Ward2018}. 

In polymer science specifically, ML 
has been used to optimise polymer-gel screening for injection wells \citep{Aldh2017} and solar cells \citep{Jorg2018}, to improve polymeric interfacial compatibilisation \citep{Meen2017}, and to classify and predict the physical features of polymer systems. For instance, supervised feed-forward neural networks have been used to recognise configurations produced from Monte Carlo simulations of polymer models, distinguishing between differently ordered states \citep{Wei2017}. Self-folding mechanisms of polymer composite systems have also been modelled \citep{Guo2013}. There is however a knowledge gap in the use of ML as a tool for classifying and predicting polymer blend morphology. ML techniques are well suited to this end as they are able to learn complex features from the data, which facilitates pattern recognition and dimensionality reduction \citep{Jie2018}. 

Previous work in the broader material science field has typically considered only one aspect of the pipeline such as employing dimensionality reduction to identify important features that contribute to the outer structure of nanoparticles \citep{Lope2012} or applying supervised machine learning for classification of new carbon black samples \citep{Mart2017}. 
The present study hence applies a machine learning workflow, comprising the use of dimensionality reduction techniques with a clustering algorithm, to separate morphological data into clusters of distinct morphologies. Pattern prediction and design-space mapping are investigated via classification, using Gaussian process classification (GPC) techniques, in the low-dimensional transformed feature space. The present analysis is restricted to PPP systems and does not consider the effects of convective mass transfer. 
The value of the workflow developed in this study is twofold: (i) the approach is generalisable to additional engineering fields beyond polymer science; and (ii) polymer blends can be studied more expediently as regions of interest (input parameters/morphology) can be first identified which allows resources (computational/experimental) to be focused.

\section{Theory and Methodology}

\begin{figure}[htbp]
	\centering
	\includegraphics[width=0.5\linewidth]{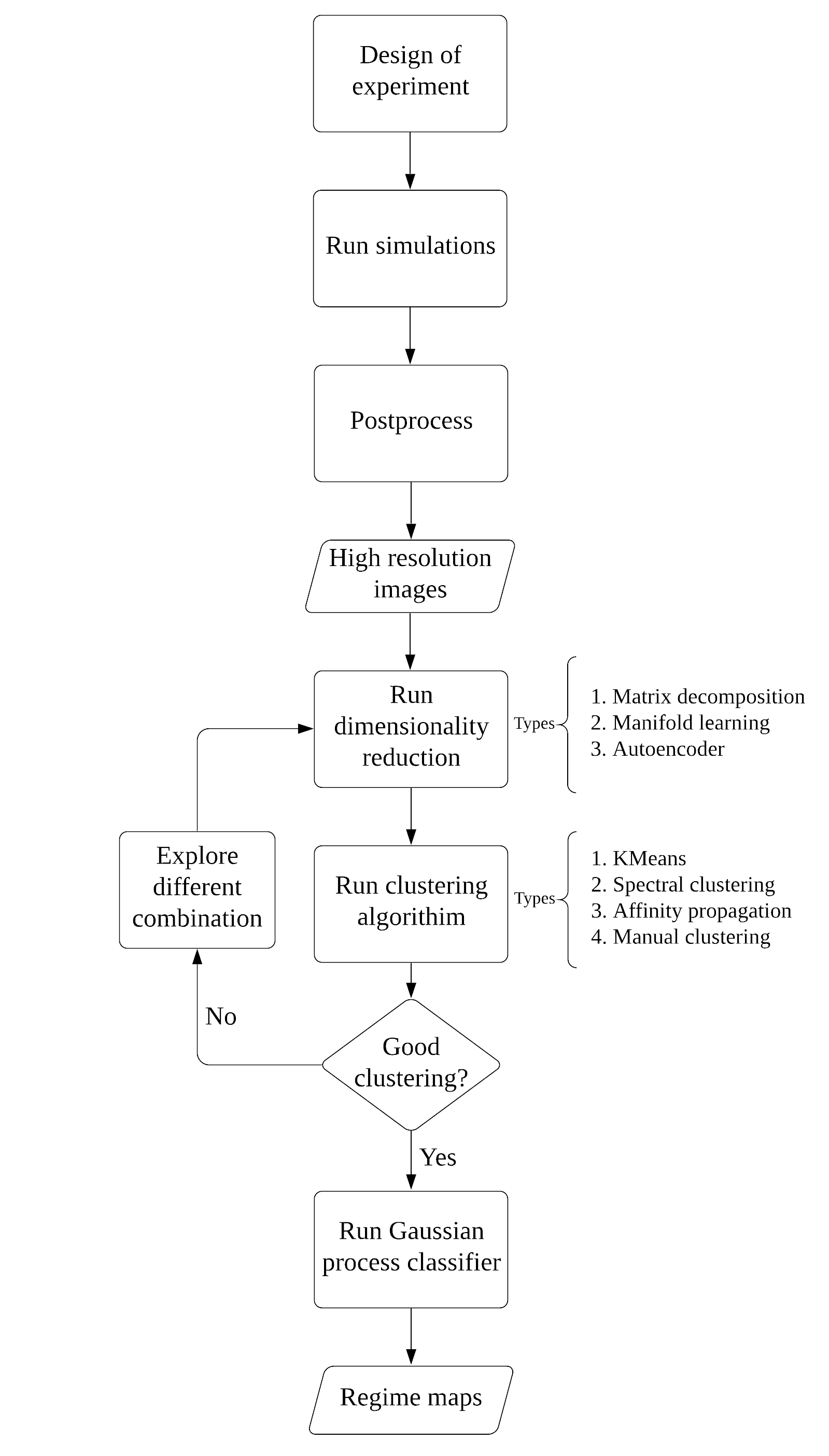}
	\caption{Workflow for integrating the physics-based simulation set with ML dimensionality reduction, clustering and prediction algorithms}
	\label{fig:workflow}
\end{figure}

We address the problem set where physics-based simulation outputs need to be predicted from known input parameters: a workflow for integrating physics-based simulations and ML clustering is outlined in \cref{fig:workflow}.

\subsection{Physics-based Cahn–Hilliard system}

The Cahn–Hilliard equation is well suited for modelling polymer blend precipitation at continuum length and time scales. Three different polymer species $a, b, c$ are tracked, however the following derivation is kept as general as possible to demonstrate the applicability of the model to $n$ component mixtures. We use a modified Cahn–Hilliard system based on the work of \citep{Petr2012} which has the advantage of being able to handle mixtures where the components may have orders of magnitude difference in diffusivities. The ability of the model to handle components with large differences in diffusivity is important in modelling systems with polymers and solvents or polymers of significantly different chain lengths \citep{Alfa2007}. 

The model used in the present work is also easier to implement than the earlier method of \citet{Alfa2007} for handling components with large difference in diffusivity. \citet{Alfa2007} introduced a `proportional flux method' within a finite difference scheme to ensure that the sum of fluxes into a point is zero. The proportional flux method introduces two issues: (i) the method itself is an heuristic solution without a robust theoretical foundation \citep{Naum2001a}, and (ii) many modern partial differential equation (PDE) solvers, for example FEniCS \citep{Fenics2011} and FiPy \citep{Guyer2009}, are primarily declarative and {\it ad hoc} adjustments to standard discretisations are difficult to implement.

The Cahn–Hilliard equation, as previously mentioned, models uphill diffusion where the driving force is gradients in the chemical potential $\mu$ rather than 
concentration. Correspondingly, the flux $\boldsymbol{j}_{i}$ of species $i$ can be represented as
\begin{equation}
	\boldsymbol{j}_{i}= - \sum_{j}L_{ij}\nabla\mu_{j} ,
\end{equation}
where each $L_{ij}$ is a species mobility coefficient and $\boldsymbol{L}$ is square symmetric. 
The following constraints are imposed on the flux expression due to the Onsanger reciprocal relations and that the total flux into a point is zero \citep{Petr2012}:
\begin{equation}
	L_{ij} = L_{ji},\quad \sum_{i}L_{ij} = 0,\quad  \sum_{i} \boldsymbol{j}_{i} = \boldsymbol{0}.
\end{equation}
Correspondingly as per \citet{Petr2012}, $\boldsymbol{j}_{i}$ can be expressed in terms of  differences in chemical potentials:
\begin{equation}
    \label{eq:fluxdiffmu}
	\boldsymbol{j}_{i}= \sum_{j}L_{ij}\nabla(\mu_{i} - \mu_{j}).
\end{equation}
As the Gibbs energy functional is scaled by $RT$, $L_{ij}$ can be expressed with the following relationship:
\begin{equation}
    \label{eq:mobility}
	L_{ij} = -D_{ij}x_{i}x_{j} ,
\end{equation}
where $x_{i}$ and $x_{j}$ are the mole fractions of species $i$ and $j$, respectively. The mobility coefficients $L_{ij}$ are composition dependent, but the $D_{ij}$ effective diffusion coefficients can be constant.

To obtain the relevant transport equation for each species, we apply the continuity equation:
\begin{equation}
    \label{eq:continuity}
	\frac{\partial x_{i}}{\partial t} + \nabla \cdot \boldsymbol{j}_{i} = 0,
\end{equation}
where $t$ is time.
To determine an expression for the chemical potential, a generalised Landau–Ginzburg free energy functional for $N$ components, $G_\text{system}$, which accounts for inhomogeneity in the system is first considered \citep{Cahn1958,Naum1989}:
\begin{align}
    \label{eq:gibbs}
   \frac{G_{\text{system}}}{RT}= \int_{V}
    \left[
        g(x_{1},x_{2},...,x_{N})  
        + \sum_{i}^{N-1}\frac{\kappa_{i}}{2}(\nabla x_{i})^2
        + \sum_{j>i}\sum_{i}^{N-1}\kappa_{ij}(\nabla x_{i})(\nabla x_{j}) 
    \right]
    \diff V, 
\end{align}
where $g$ is the homogeneous free energy contribution, and $\kappa_{i}$ and $\kappa_{ij}$ are the self and cross-gradient energy parameters, respectively. To evaluate whether the simulation has reached an equilibrium state, the Gibbs free energy was determined at each time-step as per \cref{eq:gibbs}. For polymeric systems, the homogeneous free energy is well represented by the Flory–Huggins equation:
\begin{align}
	\frac{g(x_{1},x_{2},\ldots,x_{N})}{RT}=  \sum_{i}^{N}\frac{x_{i}}{n_{i}}\ln{x_{i}}    + \sum_{j>i}\sum_{i}^{N-1}\chi_{ij}x_{i}x_{j},
\end{align}
where $n_{i}$ is the polymer chain length and $\chi_{ij}$ is the Flory–Huggins binary interaction parameter. The generalised chemical potential, applicable for inhomogenous systems, for each species $i$ can be expressed as the variational derivative of the Gibbs energy functional \citep{Cogs2010,Naum1989}:
\begin{equation}
	\mu_{i}= \frac{\delta G_{\text{system}}}{\delta x_{i}}= \frac{\partial G}{\partial x_{i}} - \nabla \cdot \frac{\partial G}{\partial \nabla x_{i}}.
\end{equation}

We replace $x_{i}$ with $a, b ,c$ to represent the mole fractions of the three species of interest: A, B and C, respectively. We can write the following equations for the differences in chemical potentials: 
\begin{align}
    \label{eq:muAB}
	\mu_{\text{AB}} = \mu_{\text{A}} -\mu_{\text{B}} = \frac{\partial g}{\partial a} - \frac{\partial g}{\partial b}  -(\kappa_{\text{A}}  - \kappa_{\text{AB}})\nabla^{2}a   +  (\kappa_{\text{B}}  -\kappa_{\text{AB}})\nabla^{2}b 
\end{align}
\begin{equation}
	\mu_{\text{AC}} = \mu_{\text{A}} -\mu_{\text{C}} = \frac{\partial g}{\partial a} - \frac{\partial g}{\partial c}-\kappa_{\text{A}}\nabla^{2}a - \kappa_{\text{AB}}\nabla^{2}b
	\label{eq:muAC}
\end{equation}
\begin{equation}
	\mu_{\text{BC}} = \mu_{\text{B}} -\mu_{\text{C}} = \frac{\partial g}{\partial b} - \frac{\partial g}{\partial c}-\kappa_{\text{B}}\nabla^{2}b - \kappa_{\text{AB}}\nabla^{2}a.
	\label{eq:muBC}
\end{equation}
The gradient energy parameters for the PPP system \citep{Naum1994} can be evaluated as follows. We specifically consider the case of all polymer species having the same radius of gyration $R_{\text{G}}$ and diffusivity:
\begin{equation}
	\kappa_{\text{A}}= \frac{2}{3}R_{\text{G}}^{2}\chi_{\text{AC}},
\end{equation}
\begin{equation}
	\kappa_{\text{B}}= \frac{2}{3}R_{\text{G}}^{2}\chi_{\text{BC}},
\end{equation}
\begin{equation}
	\kappa_{\text{AB}}= \frac{1}{3}R_{\text{G}}^{2}\bigg(\chi_{\text{AC}} +\chi_{\text{BC}} -\chi_{\text{AB}}\bigg).
\end{equation}
The compositional dependence of $\kappa_{i}$ and $\kappa_{ij}$ is neglected following an approach commonly used by similar simulation studies \citep{Alfa2007,Zhou2006,Naum1994}. This also simplifies the computations. 

Combining the expressions for the chemical potentials \crefrange{eq:muAB}{eq:muBC}, species flux \crefrange{eq:fluxdiffmu}{eq:mobility} and the continuity equation \cref{eq:continuity}, we arrive at the following transport equations tracking species A and B: 
\begin{equation}
	\frac{\partial a}{\partial t}= \nabla \cdot\bigg(D_{\text{AB}}ab\nabla\mu_{\text{AB}} + D_{\text{AC}}ac\nabla\mu_{\text{AC}}\bigg),
	\label{eq:transport_a}
\end{equation}

\begin{equation}
	\frac{\partial b}{\partial t}= \nabla \cdot\bigg(-D_{\text{AB}}ab\nabla\mu_{\text{AB}} + D_{\text{BC}}bc\nabla\mu_{\text{BC}}\bigg).
	\label{eq:transport_b}
\end{equation}
Species C is obtained by using a material balance constraint:
\begin{equation}
	c = 1 - a - b.
	\label{eq:mass_constraint}
\end{equation}

For symmetric PPP systems, where all species diffusivities can be assumed equal, the equation system given by \crefrange{eq:transport_a}{eq:mass_constraint} reduces to that considered by \citet{Naum1994}.

\subsubsection{Scaling} 

We introduce the following scalings: 
\begin{equation}
	\boldsymbol{x}=d_{\text{p}}\tilde{\boldsymbol{x}},
\end{equation}
\begin{equation}
	t = \frac{nd_{\text{p}}^{2}}{D_{\text{AB}}}\tilde{t},
\end{equation}
where $d_{\text{p}}$ is the characteristic length scale. The chemical potential and Gibbs energy functional are scaled by $RT$ (denoted from here on as $\tilde{\mu_{i}}$ and $\tilde{g}$, respectively). We consider the case of a symmetric PPP system i.e. all the polymer species have the same chain length and diffusivity, thus resulting in the following equation system:
\begin{equation}
    \label{eq:nondimmuAB}
	\tilde{\mu}_{\text{AB}} = \tilde{\mu}_{\text{A}} -\tilde{\mu}_{\text{B}} = \frac{\partial g}{\partial a} - \frac{\partial g}{\partial b}  -(\tilde{\kappa}_{\text{A}} - \tilde{\kappa}_{\text{AB}})\tilde{\nabla}^{2}a + (\tilde{\kappa}_{\text{B}} -\tilde{\kappa}_{\text{AB}})\tilde{\nabla}^{2}b, 
\end{equation}
\begin{equation}
	\tilde{\mu}_{\text{AC}} = \tilde{\mu}_{\text{A}} -\tilde{\mu}_{\text{C}} = \frac{\partial g}{\partial a} - \frac{\partial g}{\partial c}-\tilde{\kappa}_{\text{A}}\tilde{\nabla}^{2}a - \kappa_{\text{AB}}\tilde{\nabla}^{2}b,
	\label{eq:nondimmuAC}
\end{equation}
\begin{equation}
	\tilde{\mu}_{\text{BC}} = \tilde{\mu}_{\text{B}} -\tilde{\mu}_{\text{C}} = \frac{\partial g}{\partial b} - \frac{\partial g}{\partial c}-\tilde{\kappa}_{\text{B}}\tilde{\nabla}^{2}b - \tilde{\kappa}_{\text{AB}}\tilde{\nabla}^{2}a,
	\label{eq:nondimmuBC}
\end{equation}
\begin{equation}
	\frac{\partial a}{\partial t}= \tilde{\nabla} \cdot\bigg(ab\tilde{\nabla}\tilde{\mu}_{\text{AB}} + ac\tilde{\nabla} \tilde{\mu}_{\text{AC}}\bigg),
	\label{eq:nondimtransport_a}
\end{equation}
\begin{equation}
	\frac{\partial b}{\partial t}= \tilde{\nabla} \cdot\bigg(-ab\tilde{\nabla} \tilde{\mu}_{\text{AB}} + bc\tilde{\nabla}\tilde{\mu}_{\text{BC}}\bigg).
	\label{eq:nondimtransport_b}
\end{equation}

\Crefrange{eq:nondimmuAB}{eq:nondimtransport_b} model the polymer blend demixing dynamics and form the final set of equations to be solved. 

\subsubsection{Numerical implementation}

Numerical solution of the Cahn–Hilliard equation is challenging due to the fourth-order derivative: hence, the equation is typically treated as set of coupled second-order PDEs \citep{Joki2016} as shown in the equation system of \crefrange{eq:nondimmuAB}{eq:nondimtransport_b}. The system was reformulated to variational form and solved with an open-source finite-element solver, FEniCS \citep{Fenics2011}.

An unstructured square mesh of domain-length 40 dimensionless units was generated with a resolution of 80 cells along each domain boundary. Periodic boundary conditions were applied on the left and right boundaries and Neumann conditions were applied to the top and bottom domains. Unknown variables were treated implicitly and a backward Euler method was applied for time discretisation. PPP cases were simulated for a duration of $\tilde{t} = 400$ with time step $\Delta\tilde{t}= 0.02$, corresponding to a physical duration of $t=\SI{16}{s}$. 
Physical parameters were set to $R_{\text{G}} = \SI{200e-10}{m}$, $d_{\text{p}} = R_{\text{G}}$, $D_{\text{AB}} = \SI{e-11}{m^{2}s^{-1}}$. Values of $\chi_{ij}$ simulated ranged from \numrange{0.003}{0.009} via automated batch scripting \citep{DiTo2017}.
In total \num{1140} simulation runs were performed, representing a total of 5 independent input parameters. The entire simulation process took \num{\sim 80000} core-hours. Benchmarking and model validation is discussed in the supplementary material.

\subsection{Machine learning}

\subsubsection{Methodology and workflow}

The ML workflow consisted of three steps: (i) dimensionality reduction, (ii) clustering and (iii) supervised learning. Dimensionality reduction is required to address the curse of dimensionality \citep{Kriegel2009}, which can make clustering of high-dimensional data, such as raw simulation images, prohibitively expensive. Clustering on the low-dimensional processed data was used to identify morphologies. Supervised ML, using GPC, was then used to determine a relationship between the physical input parameters and the resultant morphology.

\subsubsection{Dimensionality reduction}

Each simulation generates a high-resolution colour image of the physical morphology, where RGB image channels form a proxy for species molar fraction (i.e. subject to the material balance of \cref{eq:mass_constraint}). Prior to dimensionality reduction, each images was preprocessed to $200{\times}200$ pixel resolution. The dimensionality reduction itself was applied using three candidate techniques: (i) principal component analysis (PCA), (ii) t-distributed stochastic neighbor embedding (t-SNE) and (iii) autoencoder compression \citep{Wang2016}. The three techniques used are representative of the three broad categories of techniques for dimensionality reduction available: (i) matrix decomposition or linear techniques, (ii) Manifold learning or non-linear techniques and (iii) ANN-based techniques. PCA and t-SNE were implemented using \emph{scikit-learn} \citep{Pedregosa2011}, while the autoencoder was implemented using \emph{Keras} \citep{chollet2015keras}.

For PCA and t-SNE pipelines, the set of species concentration fields was extracted as three greyscale images ($200{\times}200$); image arrays were then flattened and concatenated into a one-dimensional array of length \num{120000} ($3{\times}200{\times}200$) representing each simulation result. For the autoencoder, the preprocessed colour images were used as direct inputs.

{\it Principal component analysis (PCA)}:
For a set of input arrays, the $q$ principal components (PCs) form the orthonormal axes onto which the retained variance under projection is maximal. The PCA pipeline within \emph{scikit-learn} was undertaken using the probabilistic model of \citet{Tipp1999}: the number of retained PCs, $q$, was varied between \numlist{0;100}, corresponding to the dimensionality of the embedding.
 
{\it t-distributed stochastic neighbor embedding (t-SNE)}:
Non-linear dimensionality reduction was undertaken in \emph{scikit-learn} using the t-SNE technique \citep{Maaten2008,Wattenberg2016}. The number of embedding dimensions was varied from \numrange{2}{10}, while the perplexity was varied from \numrange{5}{50}.

{\it Autoencoder compression}:
%
An autoencoder \citep{Lecun1998} is a specific type of artificial neutral network (ANN) that consists of two sections: (i) an encoder that compresses high-dimensional data to low-dimensional `bottleneck' representation and (ii) a decoder that recovers the original data from the encoded data \citep{Hinton2006}. Autoencoders are highly suitable for dimensionality reduction, with performance often exceeding alternative techniques such as PCA \citep{Wang2016, Hinton2006}. Autoencoders are ideal for image analysis applications \citep{Chen2016}; moreover, the hidden-layer nodes at the ANN bottleneck can be exploited for downstream clustering pipelines.

Autoencoder architectures can be labelled using a convention $T$–$N$, where $T$ denotes the layer type and $N$ is the number of layers. Layer types explored presently include (i) densely connected (denoted `Dense') and (ii) convolutional (denoted `Conv'). The simplest autoencoder architectures consist of an input and output layer with one or more Dense layers in a stacked fashion: Dense–1 has a single dense encoder layer which also serves as the bottleneck, while Dense–2 and Dense–3 contain additional layers. More complex architectures, such as LeNet–5 \citep{Lecun1998}, apply a combination of Dense and Conv layers. 

All candidate autoencoders were trained on the full preprocess image data set, enabling autoencoder compression to be implemented into the same workflow, \cref{fig:workflow}, as PCA and t-SNE dimensionality reduction. Due to the small data set size, the conventional split into training and validation sets \citep{Chen2019, Petsch2017} for evaluating clustering accuracy was not undertaken.
A summary of all explored autoencoder types and hyperparameters is given in the supplementary material. 
Hyperparameter optimisation for the dense autoencoders was performed using Talos \cite{talos2019} (fractional random search over \SIrange{10}{15}{\percent} of the grid; 10 epochs), while convolutional autoencoders were tuned manually.

\subsubsection{Clustering}

To implement a purely unsupervised learning pipeline, the clustering method and/or use of an appropriate metric, heuristic or algorithm should be able to estimate the optimal number of clusters. \emph{scikit-learn} implements a variety of clustering algorithms; such as affinity propagation, spectral clustering, hierarchical clustering and $k$-means; which can be used to this end.

The popular clustering algorithm $k$-means as implemented in \emph{scikit-learn} was used to carry out clustering as a means of measuring `sign of life.' Sign-of-life in this case refers to there being initially positive results from a  naive application of a clustering algorithm which then justifies further exploration. To this end, $k$-means was found to be as effective as any of the clustering algorithms outlined above and was used as the default clustering algorithm for all the embeddings generated by the various dimensionality reduction techniques. 

The $k$-means algorithm works by clustering the data-points into $k$ groups by minimising the within-cluster sum of squares (WCSS) \citep{Yuan2019}. The algorithm requires an initialisation method (set to `k-means++') and a predetermined number of clusters, $k$. Evaluating an appropriate number for $k$ is one of the main challenges when using $k$-means and the elbow method was used. The elbow method method involves running $k$-means for a range of $k$ values and calculating the distortion or the total sum of square errors \citep{Yuan2019}. Ideally, when the number of clusters nears the real number of clusters, there is an inflexion in the distortion, showing as an `elbow' in a plot of distortions versus $k$. 

For the clustering of the t-SNE embedding of the Conv–3/4/5 autoencoder bottleneck values, $k$-means was ill suited as it is unable to effectively identify clusters where the cluster shape is non-globular. In this case, affinity propagation was used \citet{Frey2007}. Affinity propagation does not require the declaration of the number of clusters to perform clustering, but rather it performs clustering based on passing messages between the data-points which represents the fitness of one sample to exemplify the other until a set of exemplars are identified, representing the final number of clusters \citep{Frey2007}. The implementation of affinity propagation in \emph{scikit-learn} was used and it has two main hyperparameters: the preference which refers to the strength of a data-point to be an exemplar and the damping ratio which was set to 0.9 for stability. 

\subsubsection{Supervised learning}
A \emph{scikit-learn} GPC model was trained on the initial concentrations ($a_{0}$, $b_{0}$) and corresponding cluster label. A test set (\SI{20}{\percent} of the data) was used to check if the classifier returned the correct cluster, given the specific parameters. Radial basis functions (RBFs) with a length scale of $1.0$ were selected for the Gaussian process kernel. The data augmentation process is described in the results and discussion section. To perform the prediction, the initial species concentrations ($a_{0}$, $b_{0}$) were varied within a specific range ($a_0 \in [0.1,0.8]$, $b_0 \in [0.1,0.45]$) and passed into the trained GPC to evaluate the morphology maps as shown in \cref{fig:prediction}.

\section{Results and Discussion}

\subsection{Polymer demixing simulation results}

\begin{figure}[htbp!]
	\centering
	\includegraphics[width=13cm]{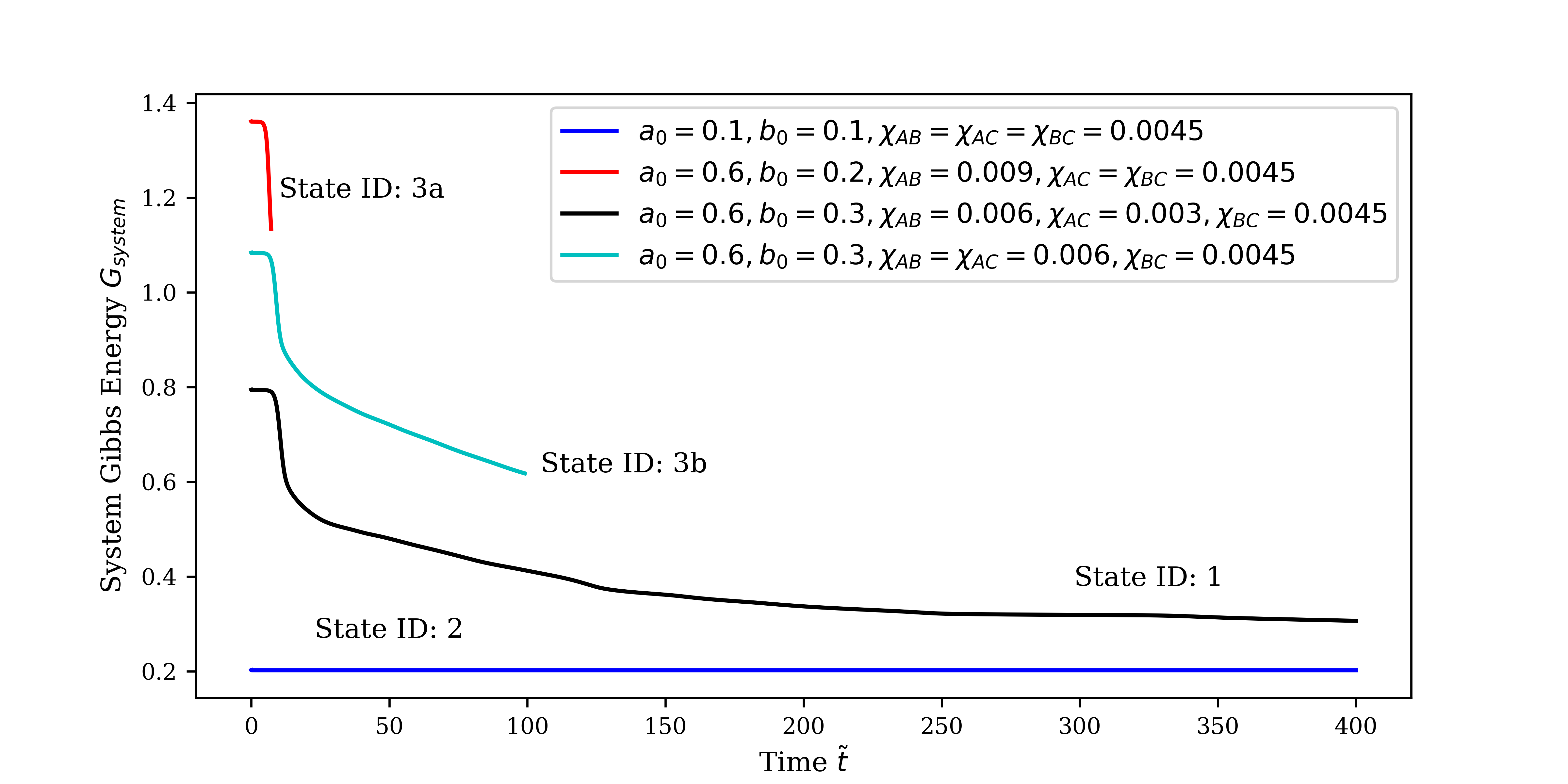}
	\caption{Representative solutions for each Gibbs energy state. The state ID for each line is annotated next to the line for reference}
	\label{fig:gibbsplot}
\end{figure}

The numerical stability of Cahn–Hilliard solution methods is a known issue, especially when using the Flory–Huggins free-energy function at higher $\chi_{ij}$ values \citep{Brun1998}. Three possible simulation states were identified as shown in \cref{tab:simstate}, demonstrating the impact of numerical issues on the solution of the physical model: either the simulation would diverge prematurely due to numerical instability (State 3a) or even though the input parameters were selected such that the physical system is in the chemical spinodal, no demixing would occur (State 2). The present data set of 629 images was restricted to samples from States 1 and 3b. A representation of how the Gibbs energy evolved with time for each of the three cases is shown in \cref{fig:gibbsplot}. 
\begin{table}[htbp!]
    \small
    \centering
    \caption{Different states that a simulation could take. Images from states 1 and 3b are used in the data set.}
    \begin{tabular}{c p{0.7\linewidth}}
    \hline
    State identity (State ID) & \multicolumn{1}{c}{Description} \\
    \hline
    1 & $G_\text{system}$ decreased over time and appeared to taper off: solutions converged for the full simulation period $\tilde{t}$ and a pattern formed. \\
    2 & Solutions converged for the full simulation period $\tilde{t}$, but $G_\text{system}$ appeared to remain constant: simulations with this State ID did not generate any patterns. \\
    3a & $G_\text{system}$ initially decreased and a pattern began to emerge, however the simulation diverged and was terminated early: the pattern was unusable. \\
    3b & $G_\text{system}$ decreased and started to plateau with the simulation generating a usable pattern, however the solution did not converge for the full simulation period $\tilde{t}$. \\
    \hline
    \end{tabular}
    \label{tab:simstate}
\end{table}

\subsection{Dimensionality reduction and clustering}

\begin{figure}[hb!]
    \centering
    \includegraphics[width = 0.9\linewidth]{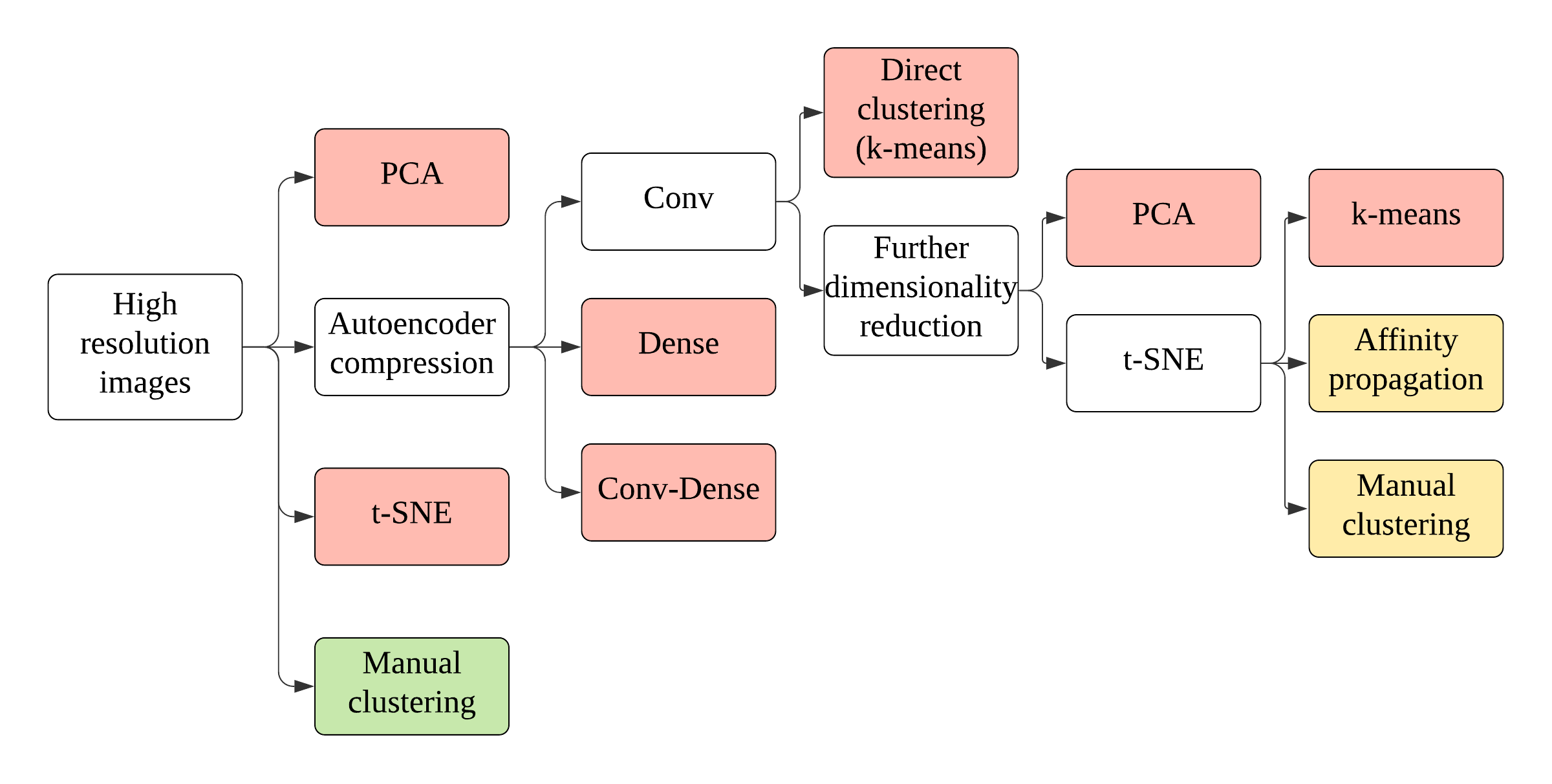}
    \caption{Dimensionality reduction and clustering results – Red: Method is unable to yield useful results; Yellow: Method is able to yield results of some significance, however the method is still inadequate; Green: Method that yielded the best results}
    \label{fig:unsupMLsum}
\end{figure}

The effectiveness of each dimensionality reduction and clustering technique is summarised in \cref{fig:unsupMLsum}: techniques were assessed.

\subsubsection{PCA}

\begin{figure}[htbp]
	\centering
	\includegraphics[width=13cm]{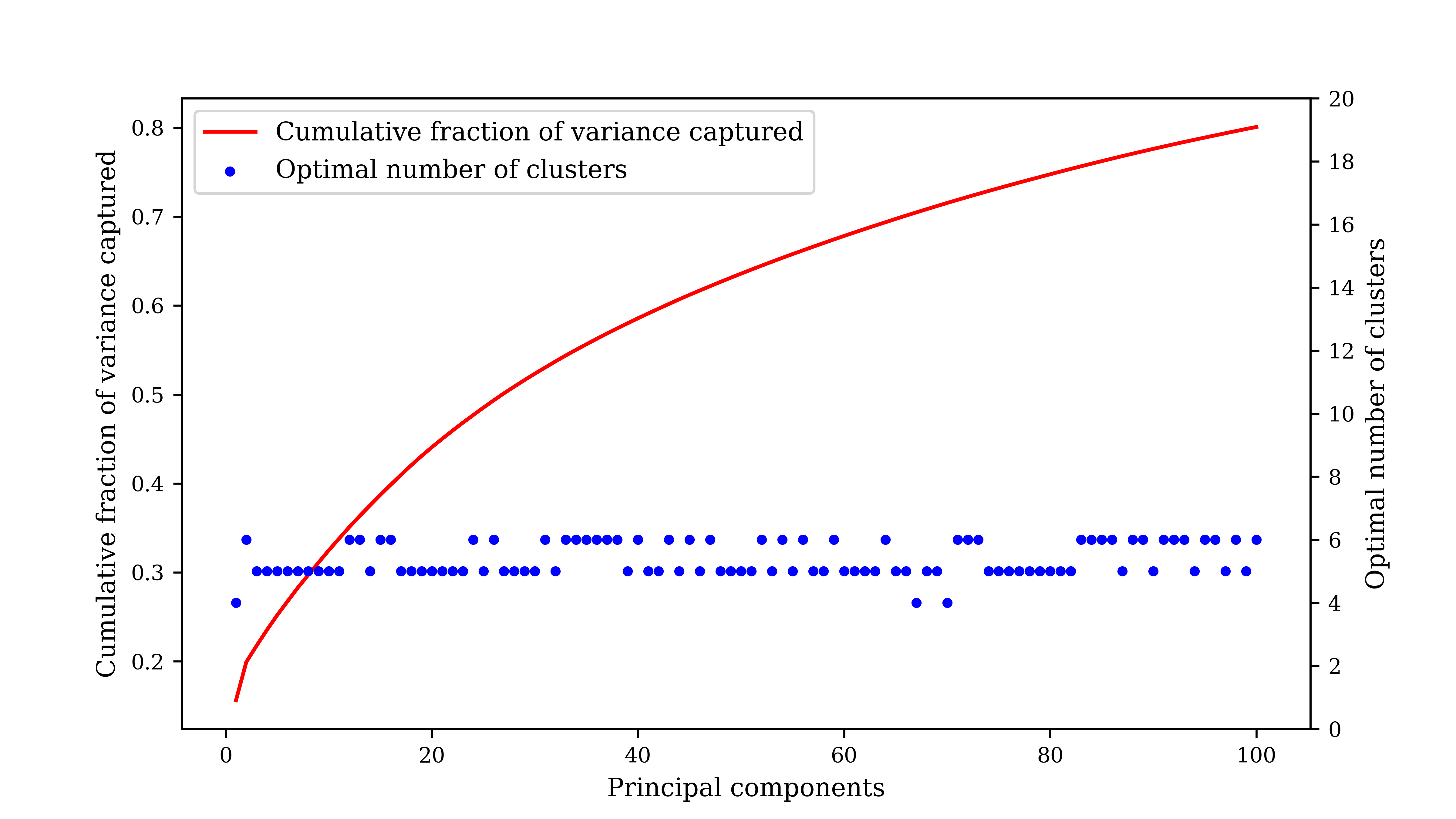}
	\caption{Captured variance and optimal cluster number: the optimal number of clusters remained between \numlist{5;6} independent of the number of PCs retained}
	\label{fig:pca_var_cluster}
\end{figure}

The image set could not by partitioned into distinct embedded-space clusters via PCA. As shown in \cref{fig:pca_var_cluster}, the number of clusters evaluated by $k$-means consistently remained between \numlist{5;6} independent of the number of retained PCs, demonstrating that the application of PCA here is ineffective. PCA was not capable of learning distinguishing features of the system as a number of clusters, each with a distinct morphology, did not emerge.

\begin{figure}[htbp]
	\centering
	\includegraphics[width=\linewidth]{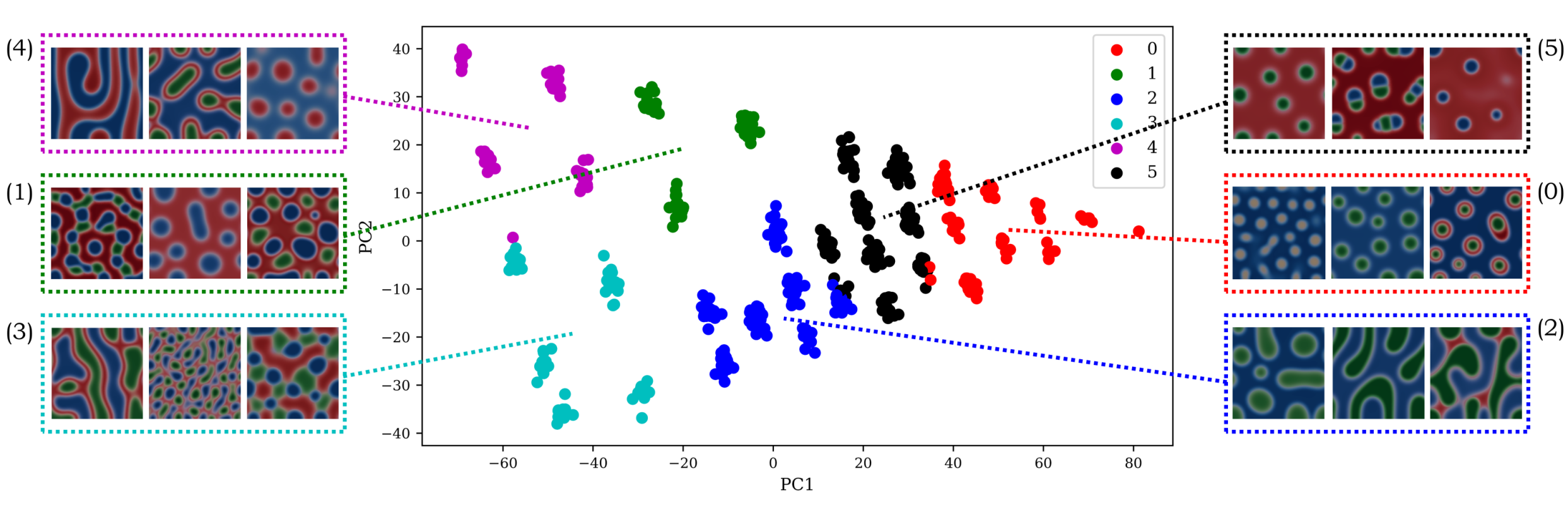}
	\caption{PCA dimensionality reduction (2 PCs retained) with $k$-means clustering (6 clusters). Sample images from each cluster are shown}
	\label{fig:2dpca_cluster}
\end{figure}

\Cref{fig:2dpca_cluster} shows how the clustering took place in 2D space: each cluster contained a variety of morphologies and was therefore not distinct. Each cluster was however consistent in terms of the predominant continuous phase. This is evidenced in the sample images for each cluster. For example, cluster 0 contains globular dispersed-phase patterns, however a mixture of core–shell (one component encapsulates the other), single-component (one component is present) and miscible (both components are mixed) patterns are present in the globular structure. A detailed description of the various observed morphologies and exemplar images is available in the supplementary material. Similar clustering behaviour was observed when retaining higher numbers of PCs.

\subsubsection{t-SNE}

\begin{figure}[htbp]
	\centering
	\includegraphics[width=0.9\linewidth]{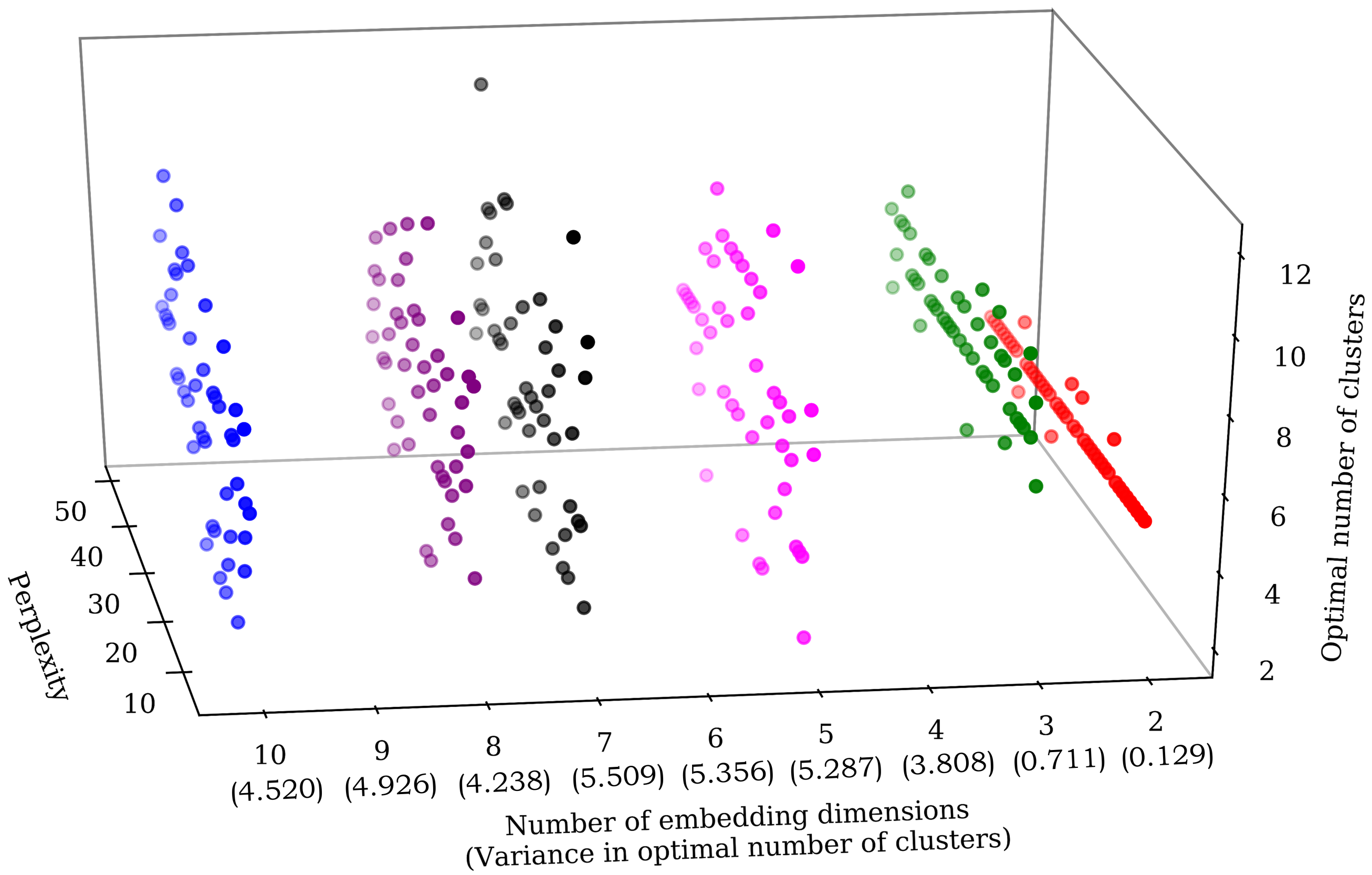}
	\caption{Number of clusters as a function of number of embedding dimensions and perplexity. Configurations with 4, 6 and 9 embedding have been omitted for clarity. The variance in the optimal number of clusters is shown in parentheses below the $x$-axis. The variance generally increases with the number of embedding dimensions}
	\label{fig:tsne_cluster}
\end{figure}

For t-SNE dimensionality reduction techniques, the variance in the optimal number of clusters was observed to increase with the number of embedding dimensions as shown in \cref{fig:tsne_cluster}. For almost all values of perplexity, the optimal number of clusters was 5 for two embedding dimensions, and \numrange{7}{8} for three embedding dimensions. A maximum in the number of clusters was observed for the combination of 7 embedding dimensions and perplexity 10.

Clustering in three or more embedding dimensions resulted in outlier clusters which contained only \numrange{1}{2} datapoints (not shown). Clusters of outlying datapoints in the embedded space distorted the $k$-means process. The optimal number of clusters is also more sensitive to the perplexity values which is reflected in the increasing variance in the number of optimal clusters. A visual inspection of the images from each cluster for sample cases revealed poorer clustering performance compared to PCA or t-SNE in two embedding dimensions with clusters having more variation in the types of morphologies and the species of the continuous phase present.

\begin{figure}[htbp]
	\centering
	\includegraphics[width=\linewidth]{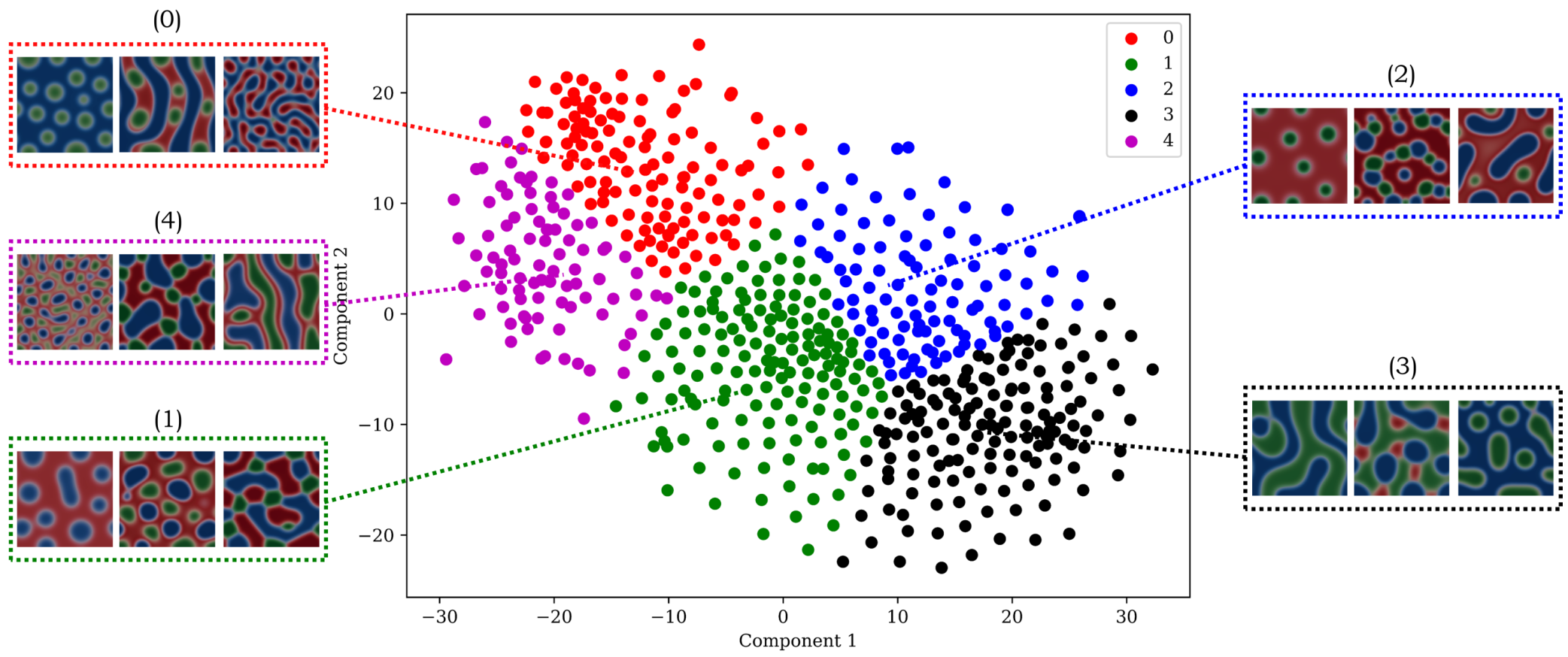}
	\caption{t-SNE results in 2D with perplexity 30 and 5 clusters. Sample images from each cluster are shown. The clustering performance is similar to the results from using PCA. There is a variety of different morphologies present in each cluster, but the species of the continuous phase is comparatively consistent}
	\label{fig:tsne_2dcluster}
\end{figure}

Use of two embedding dimensions resulted in more consistent clustering performance, with the optimal number of clusters mostly remaining at 5: a representative example can be seen in \cref{fig:tsne_2dcluster}. The dimensionality reduction and clustering performance was comparable to PCA; while there was significant mis-clustering within each cluster, each cluster was generally consistent with regards to the continuous-phase species present. Ultimately, both PCA and t-SNE techniques were unable to capture an adequate number of features to describe the diverse morphologies arising from ternary-polymer blends, prompting the exploration autoencoders as an alternative unsupervised ML workflow.

\subsubsection{Autoencoders}

The performance of each autoencoder architecture tested is shown in \cref{tab:autoencoder} together with reconstructed images and representative loss and accuracy values. Increasing the size and depth of the Dense autoencoders, which increases the number of tuneable parameters, did not improve the loss and accuracy values, which plateau at \num{\sim 0.01} and \num{\sim 0.6}, respectively, for the optimal set of hyperparameters.  The reconstructed image quality remains consistently poor. Autoencoders with Dense layers only generated poor embeddings of the images and were not further explored.

Conv autoencoder architectures performed significantly better than Dense architectures as evidenced by the reconstructed images in \cref{tab:autoencoder}. The accuracy and reconstructed image quality increased with the number of bottleneck embedding dimensions. A dimensionality of $\num{\gtrsim 500}$ is necessary to reconstruct both the morphology and continuous-phase species identity. A set of reconstructed images and loss/accuracy results for Conv–3,4,5 autoencoders is presented in the supplementary material for a range of bottleneck filter values. The addition of a Dense layer at the bottleneck of the Conv–Dense autoencoders increased the number of required parameters (\num{\geq e9}): hence, only Conv–4,5 Dense–2,1 autoencoders were tested due to their tractable memory requirements. Conv–Dense performance was comparable to Dense–1,2,3 architectures and was not further explored.

\begin{figure*}[htbp]
	\centering
	\begin{subfigure}[t!]{\linewidth}
		\centering
		\includegraphics[width=0.45\linewidth]{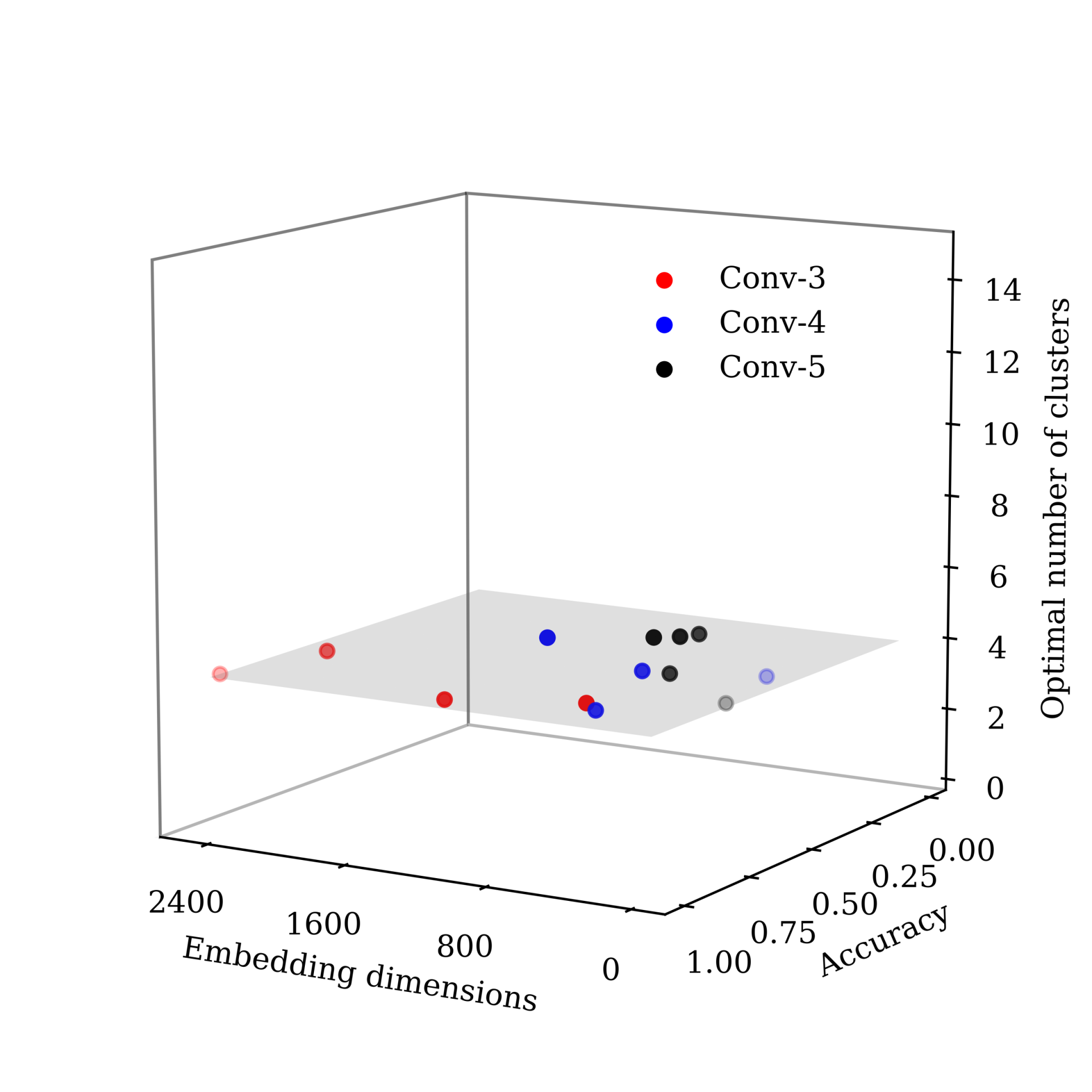}
	\end{subfigure}
	\caption{\label{fig:k-means_conv}
	$k$-means clustering on Conv–4 (4 filters) embedding (a plane of $k=4$ clusters is shown for reference). Performing $k$-means clustering directly on the Conv autoencoder bottleneck values consistently resulted in \numrange{4}{6} clusters.
	}
\end{figure*}

Clustering via $k$-means was performed directly on the Conv autoencoder bottleneck as shown in \cref{fig:k-means_conv}. Performance was found to be comparable with the other dimensionality reduction techniques tested (PCA and t-SNE): clustering on the full embedding yielded a similar number of clusters as the previous approaches. 

\begin{figure*}[htbp]
    \centering
	\begin{subfigure}[t!]{\linewidth}
		\centering
		\includegraphics[width=0.45\linewidth]{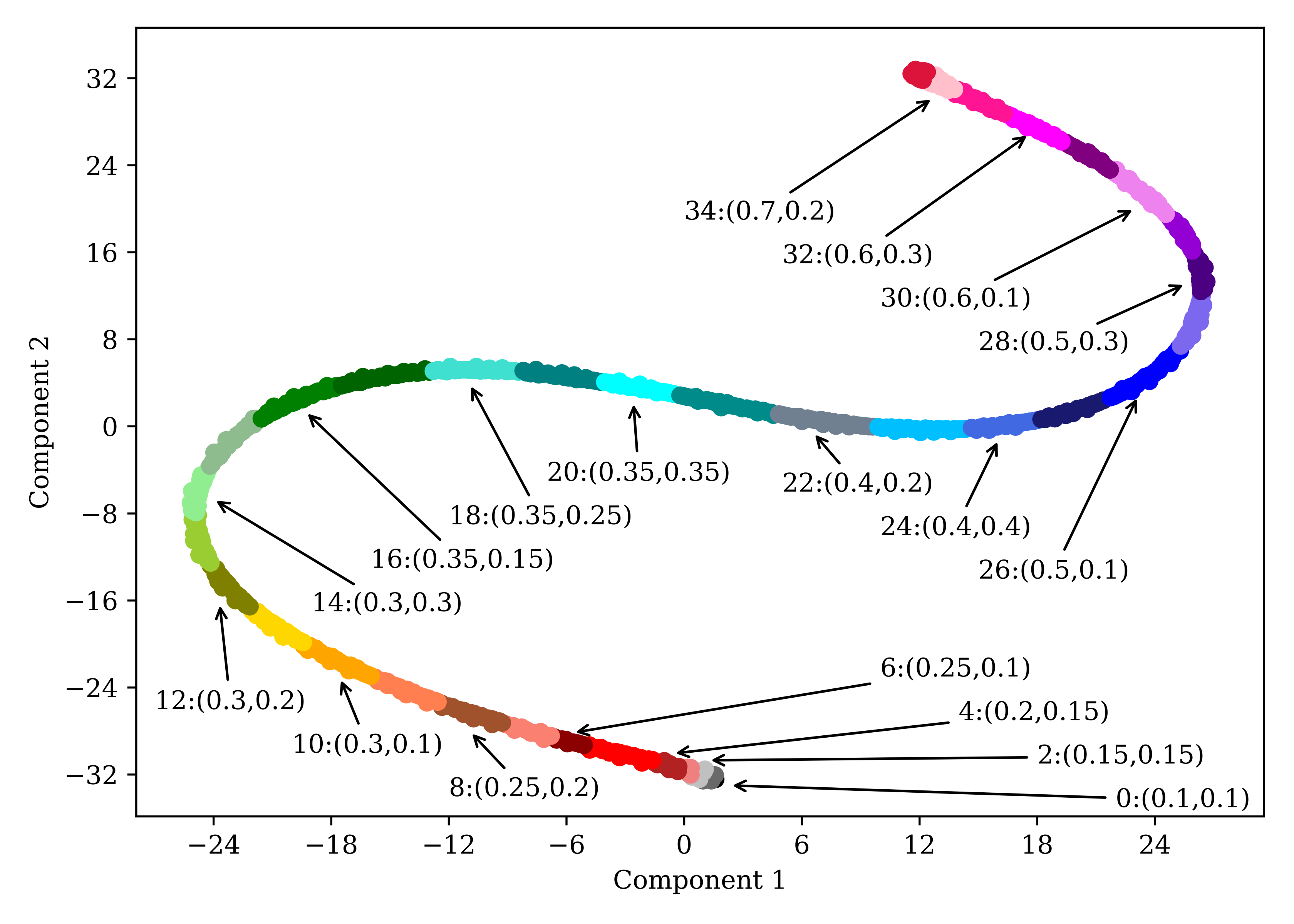}
		\caption{}
		\label{fig:concbucket}
	\end{subfigure}
	\quad
	\centering
	\begin{subfigure}[b!]{0.46\linewidth}
		\centering
		\includegraphics[width=\linewidth]{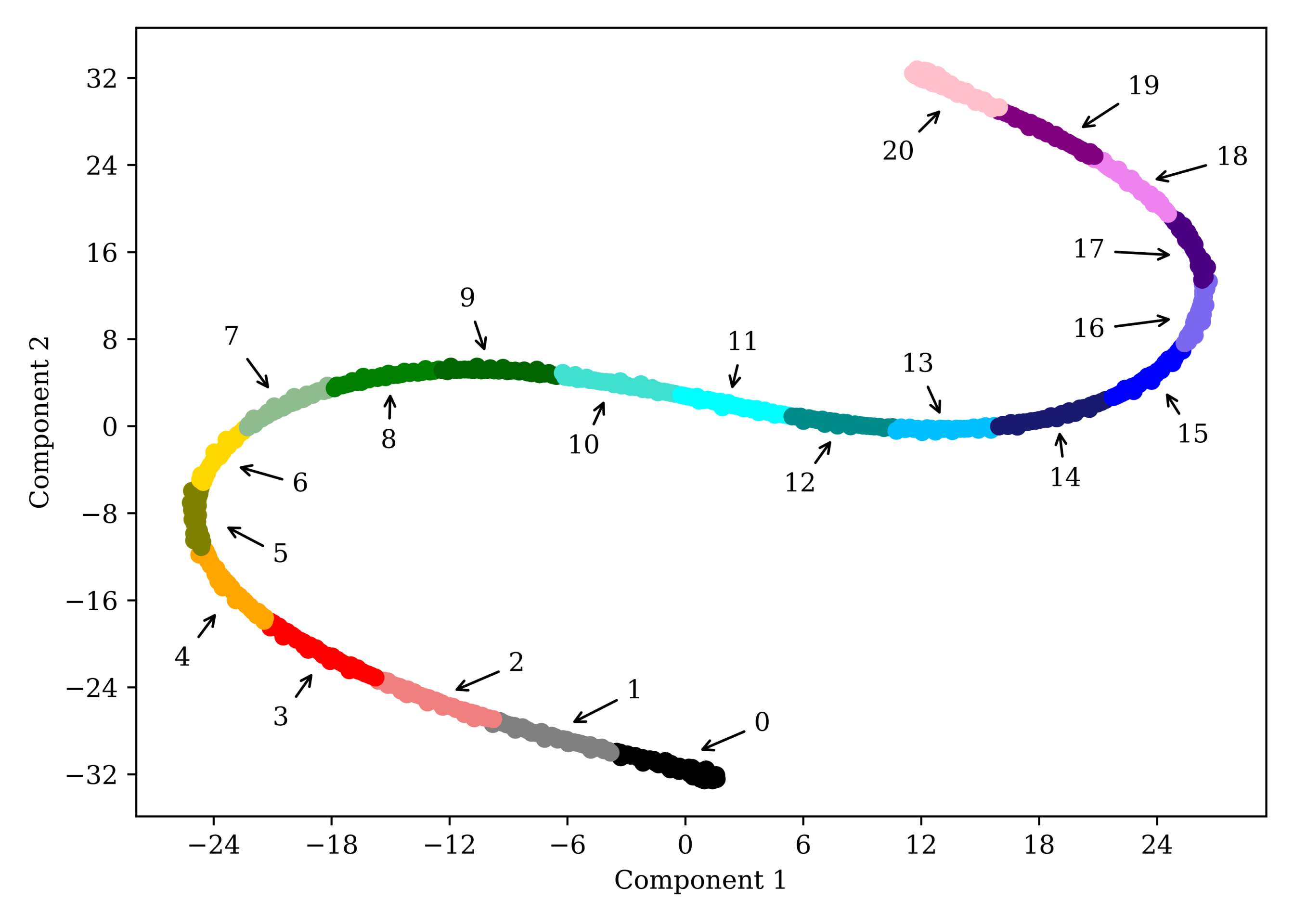}
		\caption{}
		\label{fig:affpropconvtsne}
	\end{subfigure}
	\quad
	\begin{subfigure}[b!]{0.46\linewidth}
		\centering
		\includegraphics[width=\linewidth]{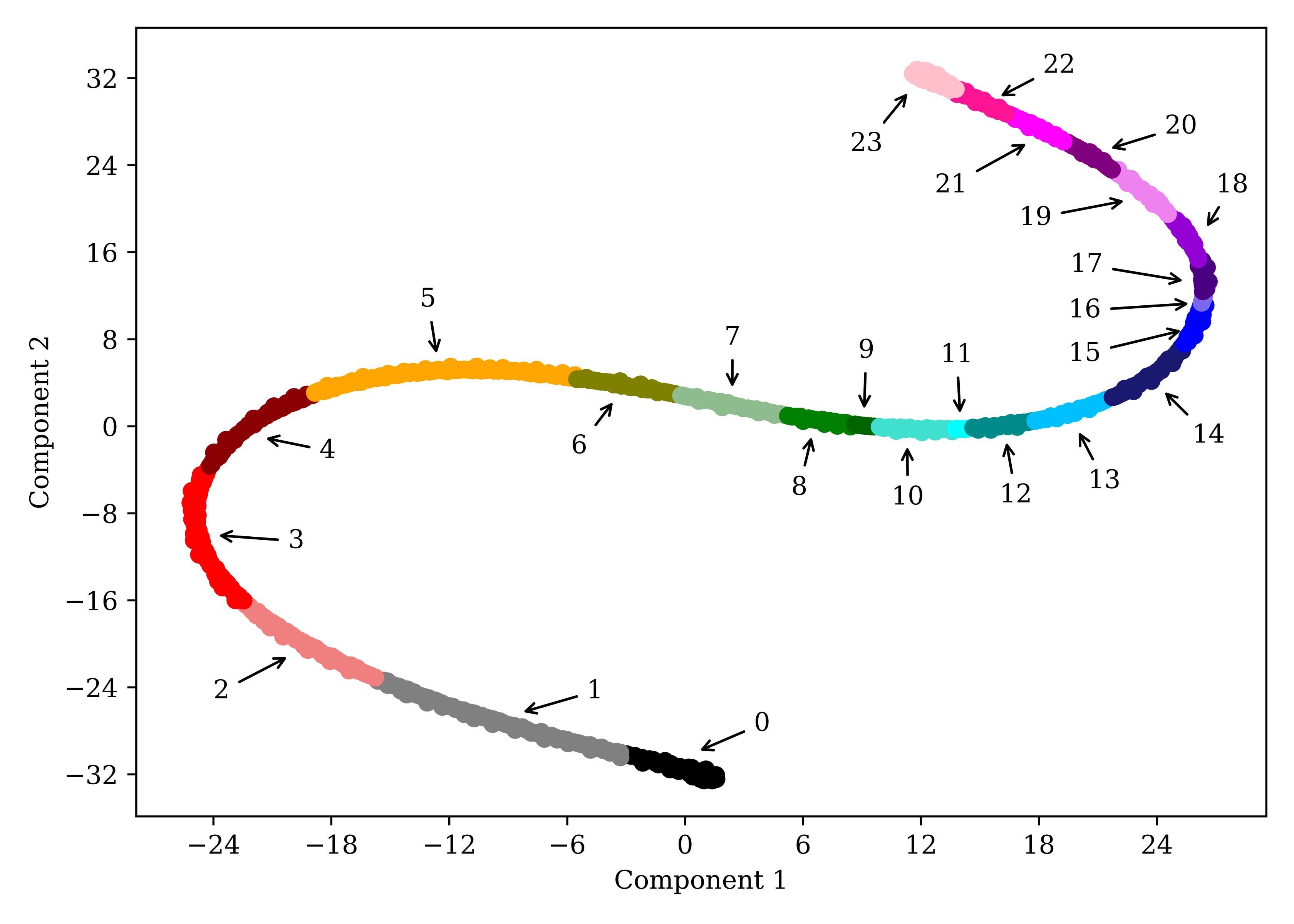}
		\caption{}
		\label{fig:manualconvtsne}
	\end{subfigure}
	\caption{%
    (a) Reduced datapoints labelled by initial composition: each group of constant $(a_{0}, b_{0})$ contains mulitple simulations with varying $\chi_{ij}$ (half of the cluster labels are omitted for clarity); (b) Affinity propagation clustering on Conv–4 (4 filters) with t-SNE embedding; (c) Manual clustering on Conv–4 (4 filters) with t-SNE embedding: applying t-SNE to the bottleneck values of Conv autoencoders arranged the datapoints based on initial composition. Affinity propagation yielded $k=21$ while manual clustering following the trend yielded $k=24$ clusters
	}%
	\label{fig:finalconvolutional}
\end{figure*}

As the dimensionality of the bottleneck was comparatively high (\numrange{\sim 100}{\sim 1000}), stacking of additional dimensionality reduction techniques to further reduce the data dimensionality was performed \citep{Maaten2008}. Applying PCA and retaining 2 PCs was not effective, yielding clustering performance comparable to applying PCA or t-SNE directly. Further dimensionality reduction was applied to the bottleneck values using t-SNE with a final two-dimensional embedding. The resulting data-point distribution shown in \cref{fig:concbucket} almost exactly coincides with the composition $a_{0}$ and $b_{0}$: by following the "S" shape curve from the bottom, the value of $a_{0}$ increases.  This result was consistent across the various Conv autoencoders and the results from the combination of the Conv–4 (4 Filters) bottleneck and t-SNE is presented in \cref{fig:finalconvolutional}. 

Clustering via $k$-means techniques is ill-suited due to the shape of the embedded data shown in \cref{fig:affpropconvtsne,fig:manualconvtsne}: the clusters cannot be ellipsoidal/spherical. Two alternative options were tested as shown in \cref{fig:unsupMLsum}: (i) manual clustering following the composition trend and (ii) affinity propagation.

Affinity propagation (with optimal preference $-250$) resulted in 24 clusters each of comparable size (\cref{fig:affpropconvtsne}), while manual clustering following the composition trend yielded 21 clusters of varying sizes (\cref{fig:manualconvtsne}).
The clustering carried out by affinity propagation and manual clustering following the composition trend had \SI{\sim 30}{\percent} of \SI{\sim 17.6}{\percent} of the datapoints within each cluster not corresponding to the majority morphology of that cluster. Furthermore, both clustering techniques resulted in non-unique clusters as different clusters would have the same majority morphology: this reduces the inherent value of each cluster as a bin to capture a distinct morphological class, adversely impacting the usability of the cluster labels for the subsequent supervised ML tasks.
The majority pattern of each cluster for both manual clustering following the composition trend and affinity propagation clustering from the Conv–4(4 Filters)—t-SNE dimensionality reduction are presented in the supplementary material together with a separate direct manual clustering of the high-resolution images based on the morphology. 

We speculate that it may be possible to obtain further improvements in the clustering performance for this case by adopting classical image-analysis and feature-engineering approaches. However, such approaches are beyond the scope of the present work as the premise of applying unsupervised ML is defeated when feature engineering is performed. Classical feature engineering and extraction identify defined features such as edges or shapes, hence classical image analysis is conceptually similar to manual clustering. In addition, while the dimensionality reduction and clustering sections of the proposed workflow had limited success in the present study, the workflow is generalisable and can be implemented with minimal modification. However, performing feature engineering constrains the workflow as it requires the features to be re-engineered for each new problem.

\begin{table}[hbt!]
\tabcolsep=4pt%
\small
\caption{Autoencoder performance for each architecture. Results for Dense–2 and Conv–Dense architectures have been omitted due to similarity with other Dense autoencoders. Dense and Conv–Dense autoencoders were observed to have lower accuracies and produce poorer image reconstructions than Conv autoencoders.}
\begin{tabular}{ccccc}
\hline
Autoencoder architecture & Encoding dimensionality & Loss / Accuracy & Sample 1 & Sample 2 \\
\hline
\addlinespace[0.1cm]
Original Images & $(200,200,3) = \num{120000}$ & 0.0 / 1.0 & 
\includegraphics[width=1.6cm,align=c]{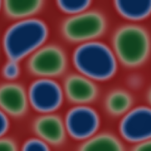}&
\includegraphics[width=1.6cm,align=c]{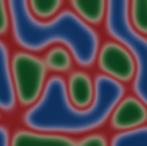} \\
\addlinespace[0.1cm]
\hline
\addlinespace[0.1cm]
Dense–1 & 500 & 0.0111 / 0.5982 & 
\includegraphics[width=1.6cm,align=c]{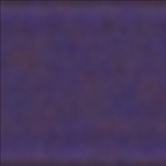}&
\includegraphics[width=1.6cm,align=c]{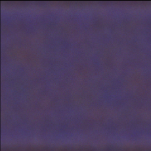} \\
\addlinespace[0.1cm]
Dense–1 & 5000 & 0.0119 / 0.6000 & 
\includegraphics[width=1.6cm,align=c]{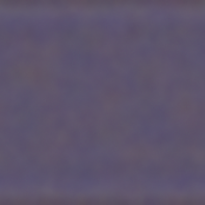}&
\includegraphics[width=1.6cm,align=c]{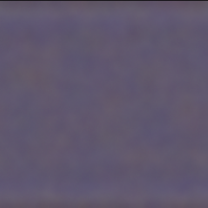} \\
\addlinespace[0.1cm]
Dense–3 & 20 & 0.0204 / 0.6039 & \includegraphics[width=1.6cm,align=c]{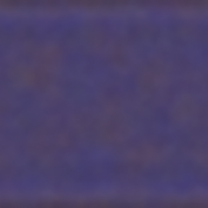}&
\includegraphics[width=1.6cm,align=c]{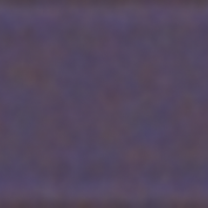} \\
\addlinespace[0.1cm]
\makecell{Conv–3 \\ (4 Filters)} & $(25,25,4) = 2500$ & $4.5042 \times 10 ^{-4} $ / 0.9654 &
\includegraphics[width=1.6cm,align=c]{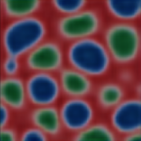} &
\includegraphics[width=1.6cm,align=c]{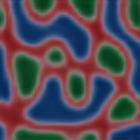} \\
\addlinespace[0.1cm]
\makecell{Conv–4 \\ (4 Filters)} & $(13,13,4) = 676$ & 0.0012 / 0.9280 & 
\includegraphics[width=1.6cm,align=c]{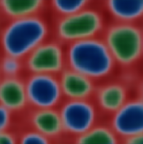} &
\includegraphics[width=1.6cm,align=c]{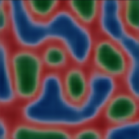}
\\
\addlinespace[0.1cm]
\makecell{Conv–5 \\ (4 Filters)} & $(7,7,4) = 196$ & 0.0050 / 0.7949 & 
\includegraphics[width=1.6cm,align=c]{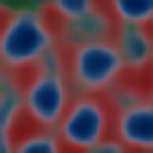} &
\includegraphics[width=1.6cm,align=c]{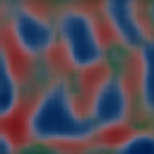}
\\

\addlinespace[0.1cm]
\hline  
\label{tab:autoencoder}
\end{tabular}
\end{table}

\subsection{Morphology prediction}

The manual cluster identities obtained by performing manual clustering directly on the high resolution images were used to train a prediction model. Even though affinity-propagation clustering and manual clustering following the composition trend on the Conv–t-SNE embedding as outlined in \cref{fig:unsupMLsum} was able to identify clusters with reasonable accuracy (\cref{fig:affpropconvtsne}), each cluster did not represent an intrinsic morphology. Therefore, the manual labels were deemed to be more appropriate for downstream supervised learning.

The simulation has a total of 8 parameters: the composition which is controlled by $a_{0}$, $b_{0}$, the polymer chain lengths $N_{i}$ for $i = 1,2,3$ and the binary interaction parameters $\chi_{ij}$ for $(i,j) = (1,2)$, $(1,3)$, and $(2,3)$. Polymers of the same chain length, $N_{i} = 1000$ for all $i$, were considered in this study which resulted in a total of 5 independent variables. For ease of visualisation, 2D slices of the 5D space are presented when we consider how the composition affects the morphology for different cases of interaction parameters. The entire image data set together with corresponding simulation parameters is available in the supplementary material.

The quantity of raw data for a given slice was approximately \numrange{25}{50} datapoints. The low count was deemed inadequate for stable predictions (whereby the prediction quality becomes independent of the data quantity), and data augmentation was hence applied. Performing additional simulation runs was not considered due to computational limitations and anticipated numerical instability issues.
Data augmentation was performed under the assumption that slight perturbations in $a_{0}$ and $b_{0}$ do not change the morphology of the polymer blend. The size of the data set was increased threefold by considering $(a_0\pm \epsilon, b_0\pm \epsilon)$ for $\epsilon \in \{0.002, 0.005\}$. The associated uncertainty introduced to the prediction was bounded by \SI{5}{\percent}. This comes about when considering the smallest datapoint $a_{0} = 0.1$, $b_{0} = 0.1$ and the largest change of $\pm 0.005$. The $\ell_{2}$ norm changes by \SI{5}{\percent} in this case and is lower for all other datapoints.  

\begin{figure*}[htbp]
	\centering
	\begin{subfigure}[t!]{0.48\linewidth}
		\centering
		\includegraphics[width=\linewidth]{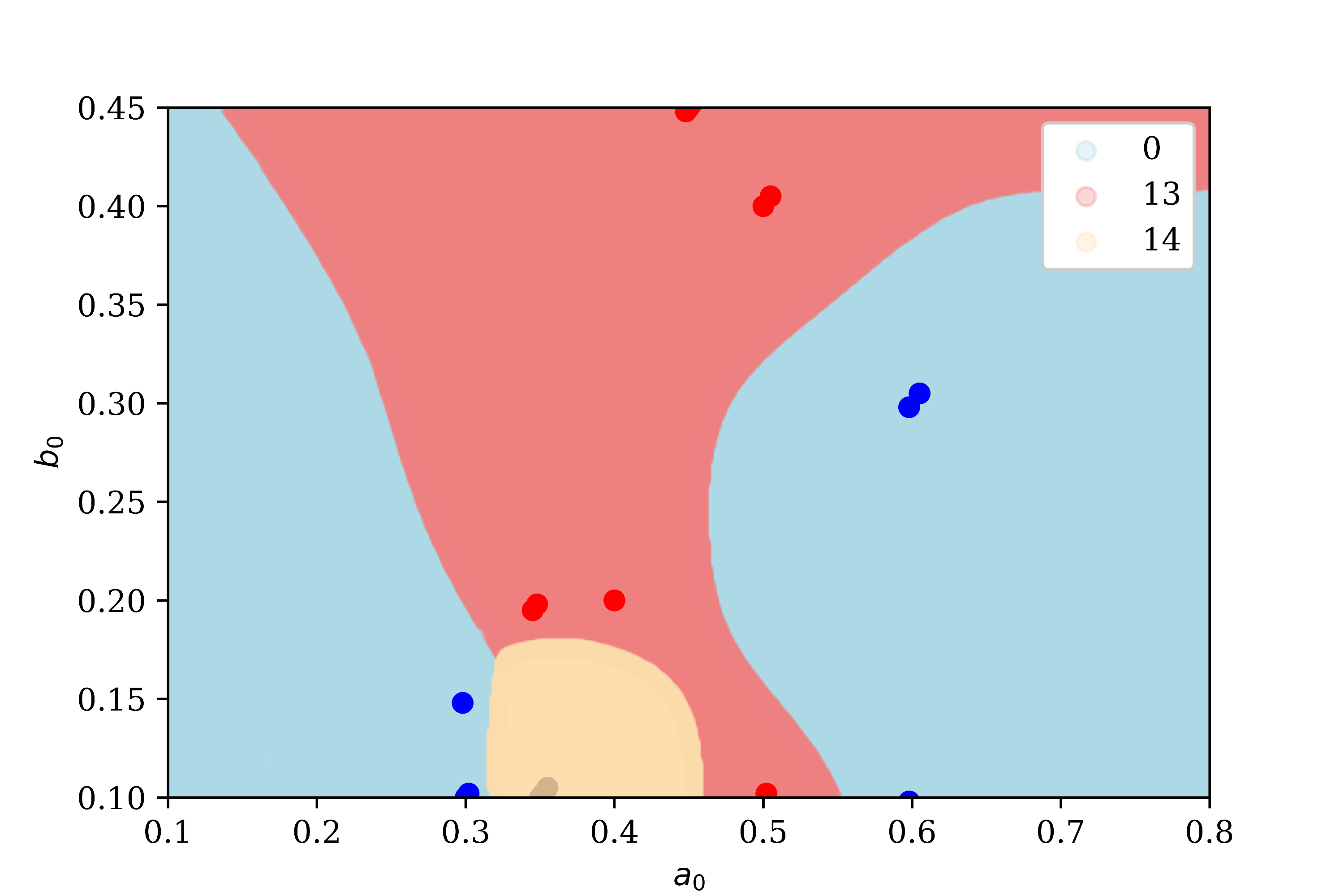}
		\caption{}
	\end{subfigure}
	\quad
	\begin{subfigure}[t!]{0.48\linewidth}
		\centering
		\includegraphics[width=\linewidth]{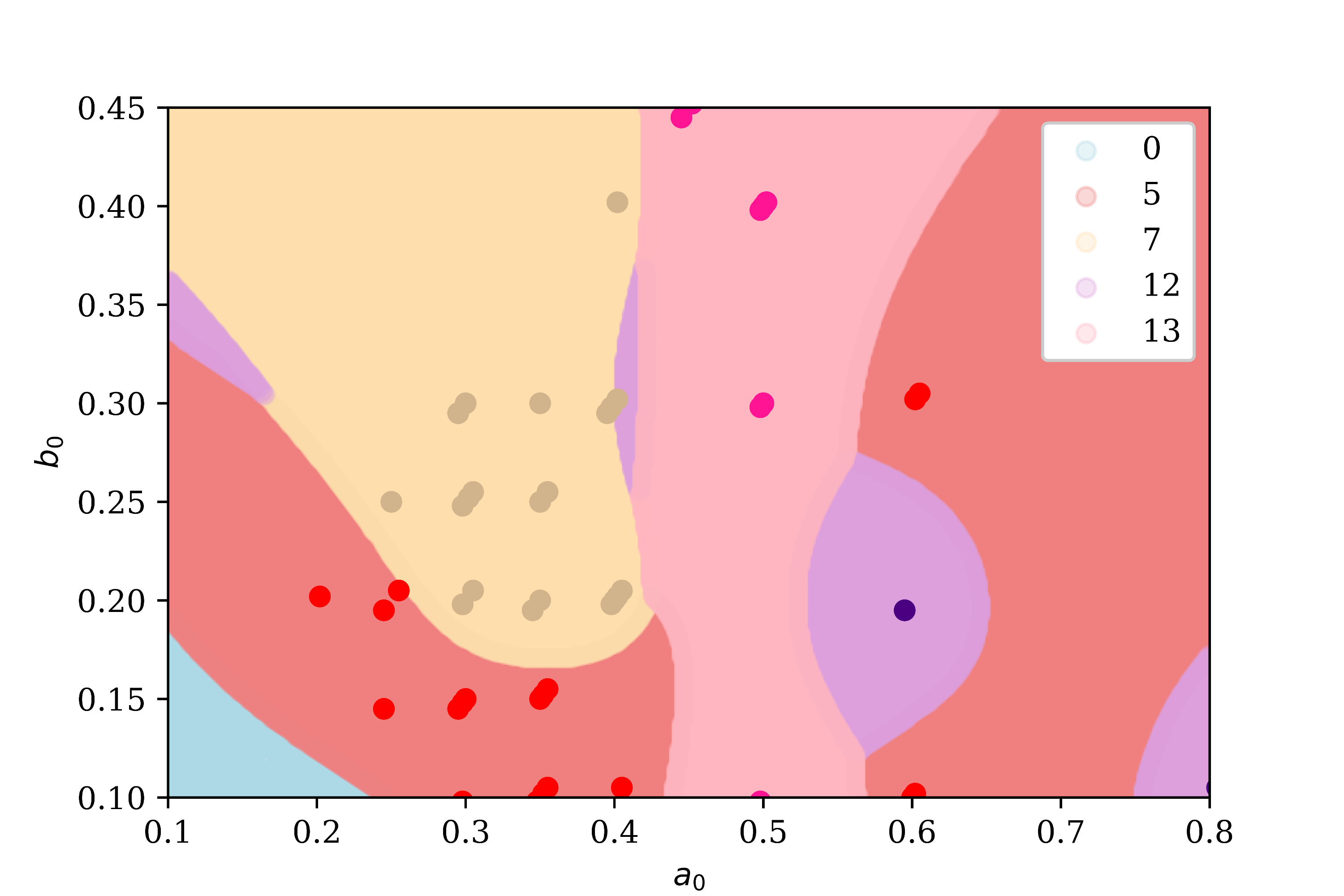}
		\caption{}
	\end{subfigure}
	\caption{Prediction of blend morphology for (a) $\chi_{ij} = \chi_{ik} = \chi_{jk} = 0.003$ and (b) $\chi_{ij} = \chi_{jk} = 0.006, \chi_{ik} = 0.003$}
	\label{fig:prediction}
\end{figure*}

As shown as \cref{fig:prediction}, the prediction process was able to yield maps whereby input values of $a_{0}$ and $b_{0}$ could be mapped to distinct morphological clusters using the cluster numbers from the manual clustering process. The test data has been overlaid onto each plot for reference. The prediction accuracy for the first case presented where $\chi_{ij} = \chi_{ik} = \chi_{jk} = 0.003$ was found to be \SI{100}{\percent}. The second case where $\chi_{ij} = \chi_{jk} = 0.006$, $\chi_{ik} = 0.003$ had a prediction accuracy of \SI{93}{\percent}. The high performance of the GPC in mapping the regimes demonstrates the capability of supervised machine learning techniques for morphology prediction.

\section{Conclusion}

Physics-based simulations of ternary polymer demixing were implemented using Cahn–Hilliard theory and a numerical solver. Despite difficulties due to numerical instabilities, a comprehensive data set was generated that covers meaningful parameter ranges, including Flory–Huggins interaction parameters and molecular size.

The performance of conventional dimensionality reduction techniques (PCA and t-SNE) when clustering the simulation results into distinct categories was inadequate for use in downstream supervised learning tasks. Application of machine learning to the present simulation set remains a challenging task: the techniques struggled to identify unique polymer-blend features that are important for morphology characterisation. It may well be possible to apply more sophisticated clustering techniques, the time and cost investments may significantly exceed those of direct manual labelling and yield comparatively poorer results. 
Supervised machine learning using GPC was used to predict the polymer blend morphology to within \SI{\geq 93}{\percent} accuracy; the accuracy is anticipated to increase with the addition of further simulation training data.

The data set (included in the supplementary material) enables users to obtain reasonable first predictions of polymer-blend morphologies for polymers with comparable physical parameters, bypassing computationally expensive simulations: resources can hence be targeted at regions of interest in the physical parameter space.
The present framework can be readily extended for ternary polymer blends with modified physical properties. Extension is also envisioned for entirely different systems, including binary polymer blends (PP) and Polymer–Polymer–Solvent (PPS) systems, and the coupling of Navier–Stokes models for prediction of shear on polymer-blend morphology.

\begin{Backmatter}


\paragraph{Funding statement}

We acknowledge Funding from the UK Research and Innovation, and Engineering and Physical Sciences Research Council through the PREdictive Modelling with QuantIfication of UncERtainty for MultiphasE Systems (PREMIERE) programme, grant number EP/T000414/1, the Alan Turing Institute AI for Science and Government programme. O.K.M. acknowledges the Royal Academy of Engineering Research Chair in Multiphase Fluid Dymamics. I.P. acknowledges funding from the Imperial College Research Fellowship scheme (ICRF).

\paragraph{Competing interests}
None

\paragraph{Data availability statement}
The full data set used in this study along with the scripts and environment configuration files needed to run the simulations and machine learning tasks can be found in the following repository: \url{https://github.com/ImperialCollegeLondon/polymer_blend_morphology}

\paragraph{Ethical standards}
The research meets all ethical guidelines, including adherence to the legal requirements of the study country.

\paragraph{Author contributions}
Conceptualisation: O.K.M; L.R.M. Methodology: L.R.M.; P.I.; I.P. Software: L.R.M.; I.P.; P.I. Investigation: L.R.M.; P.I.; M.H.; Supervision: O.K.M; L.R.M. Visualization: P.I; M.H. Writing – original draft: P.I.; M.H. Writing – review \& editing: P.I.; L.R.M.; I.P. All authors approved the final submitted draft.

\paragraph{Supplementary material}
Supplementary material intended for publication has been provided with the submission.


\bibliographystyle{apalike}

\bibliography{references,references_local}

\end{Backmatter}


\end{document}